\documentclass[12pt]{article}

\usepackage{geometry}

\usepackage{units}
\usepackage{amsbsy}
\usepackage{amstext}
\usepackage{amssymb}
\usepackage{graphicx}
\usepackage{esint}

\begin{document}

\title{Streamlines of perfect fluid as 
geodesics in Riemannian space-time}

\author{L Verozub, \\
Kharkov Karazin University\\
 leonid.v.verozub@univer.kharkov.ua}
 \maketitle
\begin{abstract}
Streamlines of a relativistic perfect isentropic fluid are geodesics
of a Riemannian space whose metric is defined by enthalpy of the fluid.
This fact simplifies the solution of some problems, as well as is
of interest from the point of view of fundamental physics.
\end{abstract}


\maketitle

\section{Introduction}

The standard method for studying ideal fluid based on solutions of
partial differential equations (Euler equations). However, at least
in the case of isentropic fluids there is another approach that is
of interest not only because of greater simplicity, but also because
it demonstrates that not only the gravity can be interpreted as the
curvature of space-time.

This approach is inspired by the existence of an effective numerical
solution of problems of hydrodynamics \cite{Monaghan1,Monaghan2},
known as Smoothed Particle Hydrodynamics (SPH). In this method, a
fluid is considered as composed by finite number of particle. These
particles move under the action of inter-particle forces which mimic
effects of pressure, viscosity, and so on. Due to the replacement
of integration by summation over number of particles, continual derivatives
become the time derivative along the particle trajectory, and as a
result, the motion of particles governed by ordinary differential
equations of classical mechanics.

\section{\noindent Equations of motion}

Consider generally accepted equations of relativistic hydrodynamics
in a manifold $\mathcal{M}$ in which defined the structure of the
Minkowski space-time $E$ with line element $d\sigma^{2}=\eta_{\alpha\beta}(x)dx^{\alpha}dx^{\beta}$,
where $\eta_{\alpha\beta}(x)$ is the metric tensor in the used coordinate
system. In inertial reference frame these equations follow from the
law of stress-energy tensor conservation\cite{Landau} :

\begin{equation}
T_{\alpha;\beta}^{\beta}=0.\label{eq:T_coservation}
\end{equation}
 In (\ref{eq:T_coservation}) the tensor $T^{\alpha\beta}=w\, u^{\alpha}u^{\beta}-p\eta^{\alpha\beta}$,
a semicolon denotes a covariant derivative in $E$ , $w=mc^{2}+\varepsilon+p$
is the enthalpy per unit volume, $mc^{2}+\varepsilon$ is the rest
internal energy, $m$ is the rest mass of particle, $p$ is the pressure
of the fluid, $c$ is the speed of light, and $u^{\alpha}=du^{\alpha}/d\sigma$.
These equations yield:

\begin{equation}
u_{\alpha}\frac{\partial(wu^{\beta})}{\partial x^{\beta}}+wu^{\beta}\frac{\partial u_{\alpha}}{\partial x^{\beta}}-\frac{\partial p}{\partial x^{\alpha}}=0.\label{eq:GydroEqs1}
\end{equation}
The expression in brackets can be written as $(nu^{\beta})(w/n)$
where $n$ is particles number density. Then, taking into account
the continuity equation $(\rho_{0}u)_{;\beta}^{\beta}=0$, were $\rho_{0}=mn$,
eqs. (\ref{eq:GydroEqs1}) can be written as 
\begin{equation}
u_{\alpha}nu^{\beta}\frac{\partial(w/n)}{\partial x^{\beta}}+wu^{\beta}\frac{\partial u_{\alpha}}{\partial x^{\beta}}-\frac{\partial p}{\partial x^{\alpha}}=0.
\end{equation}
It follows from these equations that along any world line of a fluid
element the following equations hold 
\begin{equation}
w\frac{du_{\alpha}}{d\sigma}+u_{\alpha}n\frac{d}{d\sigma}\left(\frac{w}{n}\right)-\frac{\partial p}{\partial x^{\alpha}}=0,
\end{equation}
where $d/d\sigma=u^{\beta}\partial/\partial x^{\beta}$, or more shortly
as
\begin{equation}
\frac{d}{d\sigma}\left(\frac{w}{n}u_{\alpha}\right)=\frac{1}{n}\frac{\partial p}{\partial x^{\alpha}}.\label{eq:eq_for_w/n}
\end{equation}
By using thermodynamic identity \cite{Landau} 
\begin{equation}
\frac{1}{n}dP=d\left(\frac{w}{n}\right)-Td\left(\frac{\sigma}{n}\right),\label{eq: Thermo_Identity1}
\end{equation}
 where $\sigma$ is the entropy per unit volume, equation of the motion
(\ref{eq:eq_for_w/n}) can be written as 
\begin{equation}
\frac{d}{d\sigma}\left(\frac{w}{n}u_{\alpha}\right)=\frac{\partial}{\partial x^{\alpha}}\left(\frac{w}{n}\right)-T\frac{\partial}{\partial x^{\alpha}}\left(\frac{\sigma}{n}\right).\label{eq:eq_for_w/n-2 full}
\end{equation}
This equation contains only enthalpy and entropy per particle.

If we think about fluid as of a finite collection of particles in
spirit of \cite{Monaghan1,Monaghan2}, we can interpret this equations
as describing the motion of ``particles'', having parameters of
the real gas particles along their world lines.%
\footnote{we can consider a fluid as formed by particles with mass $m$, and
$n$, satisfying the only condition $m\, n=\rho_{0}$ , where $\rho_{0}$
is the real rest fluid density.%
}

\section{Lagrangian and geometrization}

Let us show that in isentropic fluid, where $\sigma/n=Const.$, eqs.
(\ref{eq:eq_for_w/n-2 full}) are the equations of the motion of a
particle along a geodesic of a Riemannian space-time with the line
element 

\begin{equation}
ds^{2}=(G_{\alpha\beta}dx^{\alpha}dx^{\beta}),\label{ds2}
\end{equation}
where the metric tensor $G_{\alpha\beta}=\varkappa^{2}\eta_{\alpha\beta}$
, 
\begin{equation}
\varkappa=\frac{w}{\rho_{0}c^{2}}=1+\frac{\varepsilon}{\rho_{0}c^{2}}+\frac{p}{\rho_{0}c^{2}},\label{xi}
\end{equation}
 $w$ is the enthalpy per unit volume, $\rho_{0}=mn$, $n$ is the
particle number density, and $c$ is the speed of light. 

The variational principle $\delta\int ds=0$ can be written as $\delta\int Ldt=0$,
where the Lagrangian $L$ is given by \cite{VerozubFluid1}

\begin{equation}
L=-mc\left(G_{\alpha\beta}\frac{dx^{\alpha}}{d\lambda}\frac{dx^{\beta}}{d\lambda}\right)^{1/2}d\lambda,\label{Lagrangian_in_E}
\end{equation}
 where $\lambda$ is a parameter along the world line.

If to set $\lambda=\sigma$, the Lagrange equations 

\begin{equation}
\frac{d}{d\sigma}\frac{\partial L}{\partial u^{\alpha}}-\frac{\partial L}{\partial x^{\alpha}}=0\label{eq:LgrangeEqs}
\end{equation}
yield 

\begin{equation}
\frac{d}{d\sigma}\left(\varkappa u_{\alpha}\right)-\frac{\partial\varkappa}{\partial x^{\alpha}}=0,\label{eq:MyEqHydroFinal}
\end{equation}
were the condition $\eta_{\alpha\beta}u^{\alpha}u^{\beta}=1$ in the
Minkowski space-time with the signature $(+$- - - ) has been used.

Eqs. (\ref{eq:MyEqHydroFinal}) are equivalent to (\ref{eq:eq_for_w/n-2 full})
if the enthalpy per particle $\sigma/n$ is a constant which takes
place in an isentropic fluid. In this case also (\ref{eq:eq_for_w/n-2 full})

\[
\frac{\partial}{\partial x^{\alpha}}\left(\frac{w}{n}\right)=\frac{1}{n}\frac{\partial p}{\partial x^{\alpha}}.
\]
 For this reason eqs. (\ref{eq:MyEqHydroFinal}) are equivalent to

\begin{equation}
w\frac{du_{\alpha}}{d\sigma}+u_{\alpha}u^{\beta}\frac{\partial p}{\partial x^{\beta}}-\frac{\partial p}{\partial x^{\alpha}}=0.\label{MotionEquation_in_E}
\end{equation}
 Since $d/d\sigma=u^{\beta}\partial/\partial x^{\beta}$, these equation
are in agreement with the following equations for the velocity field

\begin{equation}
wu^{\alpha}\frac{\partial u_{\alpha}}{\partial x^{\alpha}}+u_{\alpha}u^{\beta}\frac{\partial p}{\partial x^{\beta}}-\frac{\partial p}{\partial x^{\alpha}}=0\label{eq:ClassicEulerEqs}
\end{equation}
 which are the general accepted form of the Euler equations. If $\sigma/n\neq Const$,
the Euler equations do not lead to eqs. (\ref{eq:MyEqHydroFinal}).
However, it is well known (\cite{Landau}) that in perfect fluid the
enthalpy per one particle preserves along its world line, i.e.

\begin{equation}
\frac{d}{ds}\left(\frac{\sigma}{n}\right)=0.\label{eq:dsdn=00003Dconst}
\end{equation}
 For this reason, the evolution of $u^{\alpha}$along the world line
giving by eqs. (\ref{eq:MyEqHydroFinal}) is correct for any perfect
fluid.

Let us consider a simple example of the advantage of using Lagrangian
(\ref{Lagrangian_in_E}). If in (\ref{Lagrangian_in_E}) to set $\lambda=t$,
(where $t$ is coordinate time), a stationary gas flow along the axis
$x$ is described by the Lagrangian
\begin{equation}
L=mc\varkappa(x)\left(c^{2}-v^{2}\right)^{1/2},\label{eq:LagrangianPlaneMotion}
\end{equation}
where $v=dx/cdt$ . From this it is easy to find that differential
equation of the motion is 

\[
\dot{v}-\frac{\varkappa'}{\varkappa}\left(1-\frac{v^{2}}{c^{2}}\right)=0.
\]
 where a point and prime denote a differentiation with respect to
$t$ and $x$, respectively.

Since $L$ does not depend on time $t$, the law of the energy $\mathcal{E}$
conservation holds, that is  $\dot{x\, dL/d\dot{x-L}}=\mathcal{E}$, which yields
the relativistic Bernoulli equation \cite{Landau}:

\[
\frac{\varkappa\, c^{2}}{(c^{2}-v^{2})^{1/2}}=Const.
\]

Thus, instead to find a velocity field with Euler's PDF equations,
we can observe the motion of some separate tiny elements of the fluid,
which gives a complete picture of the motion of the fluid.

The differential form (\ref{ds2}) defines in the manifold $\mathcal{M}$
a line element of a Riemannian space-time $V$, so that the Lagrange
equations (\ref{eq:LgrangeEqs}) with the parameter $\lambda=s$ give
the standard equations of geodesic line in $V$ : 

\begin{equation}
\frac{du^{\alpha}}{ds}+\Gamma_{\beta\gamma}^{\alpha}u^{\beta}u^{\gamma}=0,\label{eqsPart_as_Geodesic}
\end{equation}
where 

\[
\Gamma_{\beta\gamma}^{\alpha}=\frac{1}{2}G^{\alpha\delta}\left(\frac{\partial G_{\delta\beta}}{\partial x^{\gamma}}+\frac{\partial G_{\delta\gamma}}{\partial x^{\beta}}-\frac{\partial G_{\beta\gamma}}{\partial x\delta}\right).
\]

For the metric $g_{\alpha\beta}=diag(-1,-1,-1,+1)$ the Christoffel
symbols are read:

\[
\Gamma_{\beta\gamma}^{\alpha}=-\frac{1}{\phi}\left(\frac{\partial\varkappa}{\partial x^{\gamma}}\delta_{\beta}^{\alpha}+\frac{\partial\varkappa}{\partial x^{\beta}}\delta_{\gamma}^{\alpha}-\eta^{\alpha\delta}\frac{\partial\varkappa}{\partial x^{\delta}}\delta_{\beta}^{\alpha}.\right)
\]
The curvature of this space-time is other than zero. For example,
for a stationary gas flow along $x-$axis which is described by the
Lagrangian (\ref{eq:LagrangianPlaneMotion}) the scalar curvature
is given by 
\[
R=\frac{6}{\varkappa}\left(\frac{\partial^{2}\varkappa}{\partial x^{2}}-\frac{1}{c^{2}}\frac{\partial^{2}\varkappa}{\partial t^{2}}\right),
\]
which is other than zero. 

The component 
\begin{equation}
\Gamma_{00}^{1}=\frac{1}{\varkappa}\frac{\partial\varkappa}{\partial x^{1}},
\end{equation}
and in due of eqs. (\ref{eqsPart_as_Geodesic}), in non-relativistic
limit eqs. (\ref{eqsPart_as_Geodesic}) lead to the equation 
\begin{equation}
\rho\frac{d\mathbf{v}}{dt}=-\mathbf{\boldsymbol{\nabla}}P.
\end{equation}

Thus the Lagrangian $L$ describes the motion of the particles both
in $E$ and $V$ . In the first case $G_{\alpha\beta}$ is some tensor
field in $E,$ in the second case it is a fundamental tensor of the
Riemannian space-time $V.$

Space-time $V$ is real physical space-time because an observer in
co-moving reference frame of the fluid can observe deviation of geodesic
lines exactly as in gravitational field due the fact that the space-time
curvature is other than zero. This means that at least isentropic
fluid can be considered not only by conventional manner but also as
a manifestation of curvature of space-time with a metric defined by
the enthalpy.

\section{Fluid in gravitational field}

There are two ways to consider an ideal fluid in a gravitational field,
from the above point of view.

First, the change of the fluid energy $\mathcal{E}$ in in gravitational
field is $d\mathcal{E}=Td\Sigma-pdV+d\mathcal{\mathit{\mathcal{E}}_{\textrm{gr }}}$
, where$T$ is the temperature, $\Sigma$ is the entropy, $V$ is
a volume and the last term is the change of the gravitational potential
energy. For this reason the change of the enthalpy $w$ per one particle
is given by

\[
d\left(\frac{w}{n}\right)=Td\left(\frac{\sigma}{n}\right)+\frac{1}{n}dp+d\left(\frac{e}{n}\right),
\]
 where $e$ is the density of gravitational energy. 

According to this, at the presence of gravitational field we should
set in the Lagrangian (\ref{Lagrangian_in_E}) $G_{\alpha\beta}=\varkappa^{2}\eta_{\alpha\beta}$,
where

\[
\varkappa=\frac{w}{nmc^{2}}=1+\frac{\epsilon}{\rho_{0}c^{2}}+\frac{p}{\rho_{0}c^{2}}-\frac{U}{c^{2}},
\]
 and $U$ is the gravitational potential. Now in eq. (17) appears
an additional term $\nabla U/c^{2}$. Consequently, in the equilibrium
state the ordinary condition $\rho\nabla U=\nabla p$ holds.

In relativistic case it is easy to use more traditional geometrical
approach. The presence of gravity can be accounted by considering
the line element (1) in a Riemannian rather than pseudo-Euclidean
space-time, so that the metric tensor at the presence gravity is given
by
\begin{equation}
G_{\alpha\beta}=\varkappa^{2}g_{\alpha\beta},\label{Ref: GalhabetaGeneral}
\end{equation}
where $g_{\alpha\beta}(x)$ is the metric tensor of space-time at
presence gravity.

That is, each small element of a relativistic perfect isentropic fluid
moves along a geodesic of the Riemannian space whose line element
is (\ref{Ref: GalhabetaGeneral}).

The equation of the motion of the fluid element are of the form of
standard geodesic equations in the Riemannian space-time with the
metric tensor $G_{\alpha\beta}$. If we are located very far from
the source of gravity, where space-time is, in fact, Minkowskian,
it is naturally to use the timing coordinate $t=x^{0}/c$ as a parameter
along the line, and the equations take the form

\begin{equation}
\ddot{x}^{\alpha}+\left(\Gamma_{\beta\gamma}^{\alpha}-c^{-1}\Gamma_{\beta\gamma}^{0}\dot{x}^{0}\right)\dot{x}^{\beta}\dot{x}^{\gamma}=0\label{eq:GeodesicEquations_t}
\end{equation}

where $\dot{x}^{\alpha}=\ddot{x}^{\alpha}=dx^{\alpha}/dt$. Zero component
of these equations is satisfied identically, and rest equations are
the ones for the 3-spacial velocity.

To verify this, consider for example a fluid in a gravitational field
in state of equilibrium in a spherically-symmetric gravitational field.

Since $\dot{x}^{\alpha}=\ddot{x}^{\alpha}=dx^{\alpha}/dt$ , the conditions
of the equilibrium in spherical coordinates is
\[
\Gamma_{00}^{1}=\frac{1}{2}G\left(G_{00}\right)'=0
\]
where a prime denotes derivative with respect to the radial distance
$r.$ It means that$\left(\varkappa^{2}g_{00}\right)'=0$ or
\[
\frac{\varkappa'}{\varkappa}=-\frac{\left(g_{00}\right)'}{2\, g_{00}}.
\]
Due to the equality $\varkappa'/\varkappa=p'/\rho c^{2}\varkappa$,
and taking into account (\ref{xi}), we obtain for the fluid with
$\varepsilon=0$ the standard equation of the equilibrium in General
Relativity \cite{Weinberg}
\begin{equation}
\frac{\left(g_{00}\right)'}{g_{00}}=
-\frac{2p'}{\rho c^{2}+p}.\label{eq:EquilibriumEquationGR}
\end{equation}
Thus, the motion of fluid particles in pseudo-Euclidean space-time
are at the same time some equations of the motion of these particles
along geodesic lines of the Riemannian space-time $V$.

This fact allows us to take advantage of knowledge of the Lagrangian,
in particular in the case of the existence of symmetries. As an example,
by using the symmetry of the Lagrangian, we consider the classification
of  gas flows at stationary gas accretion onto the compact object
like a neutron star or a supermassive compact object at the center
of our Galaxy.

\section{Example }

The knowledge of the Lagrangian allows to obtain a simple classification
of gas flow at the spherically-symmetric accretion. Because we observe
the motion of separate elements of fluid, this analysis is very like
the same problem for study of the motion of test particles in relativistic
mechanics \cite{Shapiro}.

According to the state above in the previous section, we start from
the Lagrangian $L=-mc\left(G_{\alpha\beta}(x)\, d\dot{x}^{\alpha}d\dot{x}^{\beta}\right)^{1/2}$,
where $G_{\alpha\beta}=\varkappa^{2}g_{\alpha\beta}$, and $g_{\alpha\beta}(x)$
is the the classical Schwarzschild solution of Einstein's gravitation
equations, that is 

\begin{equation}
L=-mc\varkappa\left[a(r)\, c^{2}-\frac{1}{a(r)}\dot{r}^{2}-r^{2}\dot{\varphi}^{2}\right]^{1/2},\label{eq:LagrangianGR}
\end{equation}
 where $a(r)=1-r_{g}/r,$ and $r_{g}=2\gamma M/c^{2}$ is the Schwarzschild
radius ($M$ is the central mass and$\gamma$ is the gravitational
constant).

The equations of the motion of test particles in the plane $\theta=\pi$/2
can be obtained by the law of conservation of the energy $\mathcal{E}$
and the angular moment $J$, which follow from the fact that $L$
does not depend on time $t$ and $\varphi$ :
\begin{equation}
\dot{r}\frac{\partial L}{\partial\dot{r}}+\dot{\varphi}\frac{\partial L}{\partial\dot{\varphi}}-L=\mathcal{E}\label{eq:EnergyConserv}
\end{equation}
and
\begin{equation}
\frac{\partial L}{\partial\dot{\varphi}}=J.\label{eq:AngularMomentumConserv}
\end{equation}
It yields the following equations of the motion of the gas element
\begin{equation}
\dot{r}^{2}=a^{2}c^{2}\left[1-\frac{a\varkappa\, c^{2}}{\bar{\mathcal{E}}^{2}}\left(1+\frac{\bar{J}^{2}}{\varkappa\, r^{2}}\right)\right]\label{eq:EqMotionGR1}
\end{equation}
 and
\begin{equation}
\dot{\varphi}=\frac{c^{2}a\, J}{r^{2}\mathcal{E}},\label{eq:EqMotionGR2}
\end{equation}
where $\overline{\mathcal{E}}=\mathcal{E}/\rho_{0}$ and $\overline{J}=J/r_{g}\rho_{0}$.
The magnitude $\mathcal{E}=\mathcal{E}(r,\dot{r})$ is the total energy,
it includes both the ``kinetic'' and ``potential '' one. Evidently,
the value of $\Upsilon=\mathcal{E}(r,\dot{r})$ at $\dot{r}=0$ is
an effective potential energy of a gas element. It follows from (\ref{eq:EqMotionGR1})
that 
\[
\Upsilon(r)=c^{2}\left(1-\frac{r_{g}}{r}\right)\left(\varkappa(r)+\frac{\bar{J}^{2}}{r^{2}}\right).
\]
The effective potential for particles in vacuum can be obtained from
this equation by setting $\varkappa(r)=1$. 

\begin{figure}
\begin{minipage}[t]{0.45\columnwidth}%
\includegraphics[width=7cm,height=6cm]
{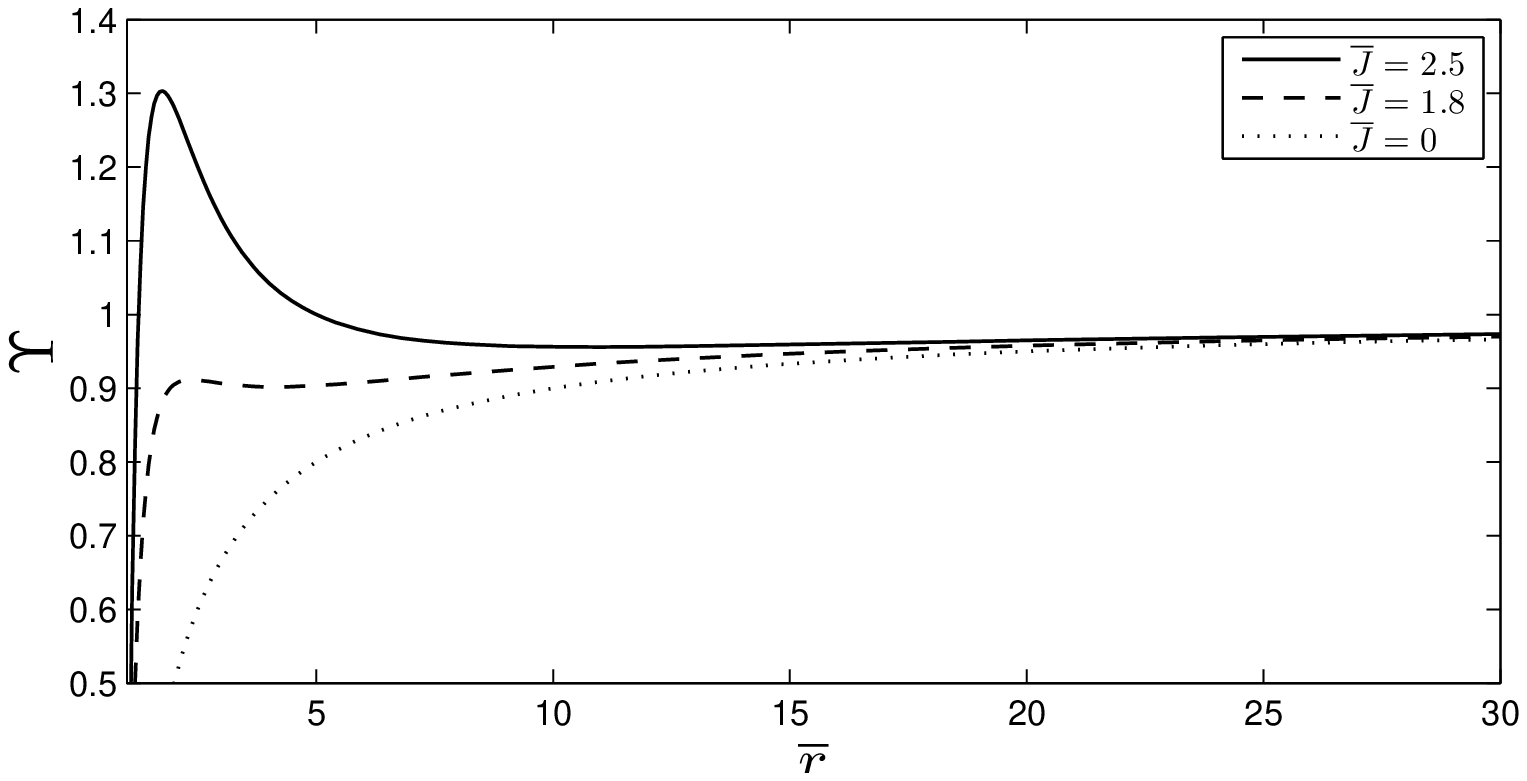}
\caption{\label{fig:EfPotParticle}The Effective potential for a free falling
test particle near the attractive point mass for $\bar{\mathcal{E}}=1$.
The potential has minimum $\bar{r}_{min}=10.7$ for $\bar{J}=2.5$
and $\bar{r}_{min}=4.1$ for $\bar{J}=1.8$.}
\end{minipage}\hfill{}%
\begin{minipage}[t]{0.45\columnwidth}%
\includegraphics[width=7cm,height=6cm]
{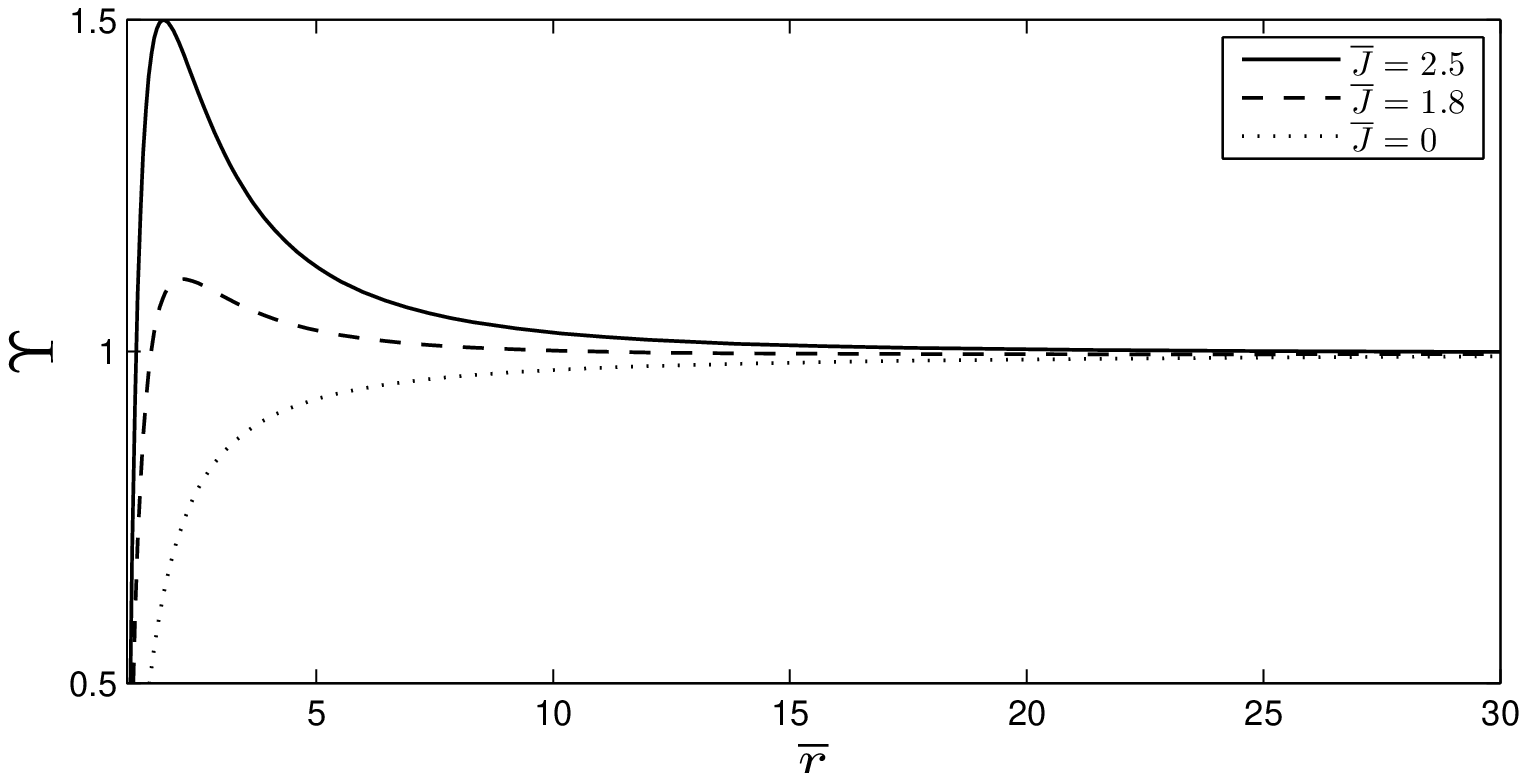}
\caption{\label{fig:EfPotGas}The Effective potential for an accretion gas
particle near the attractive point mass for $\bar{\mathcal{E}}=1$.
The potential has minimum $\bar{r}_{min}=52.7$ for $\bar{J}=2.5$
and $\bar{r}_{min}=22.2$ for $\bar{J}=1.8$.}
\end{minipage}
\end{figure}

Fig. \ref{fig:EfPotParticle} and \ref{fig:EfPotGas} show the effective
potentials for a test particle freely falling to a compact object,
and for a gas particle, respectively, for the same parameters $\bar{J}$
and $\bar{\mathcal{E}}=1$. For an illustrative purpose the gas is
supposed to be one-atomic with $\varepsilon/n=\nicefrac{3}{2}kT$
and $P=nkT$, and the temperature dependence on $\bar{r}$ 
is $\thicksim1/\bar{r}$
, so that we set $\varkappa(\bar{r})=0.8/\bar{r}.$ 

We see that similarly the case of free particles the flow gas can
have finite motion due to the existence of a potential well where
$\Upsilon(\bar{r})$ has a minimum. Positions of these minimums ($\bar{r}_{min})$
for high-temperature gas are very different from this for test particles
in vacuum. Finite gas motion (in particular, accretion disks) play
important pole in astrophysics. This method makes it easy to find
the relation between the distance of the circular motion of the gas
from the central object and physical conditions in the accreting gas.

\begin{figure}
\begin{minipage}[t]{0.46\columnwidth}%

\includegraphics[width=10.1cm,height=9cm,keepaspectratio]
{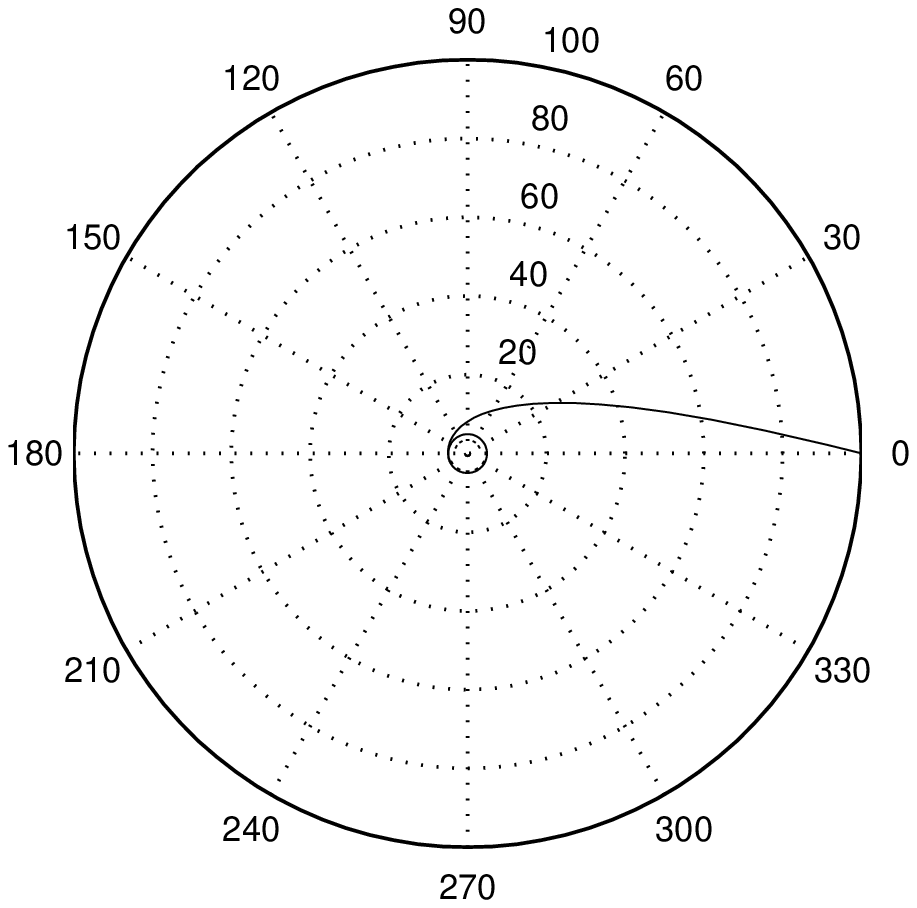}

\caption{\label{fig:gasorbitParticle}The obit of a free falling test particle
in vacuum from the distance $\bar{r}=100$. }
\end{minipage}\hfill{}%
\begin{minipage}[t]{0.46\columnwidth}%
\includegraphics[width=8.3cm,height=6.8cm,keepaspectratio]
{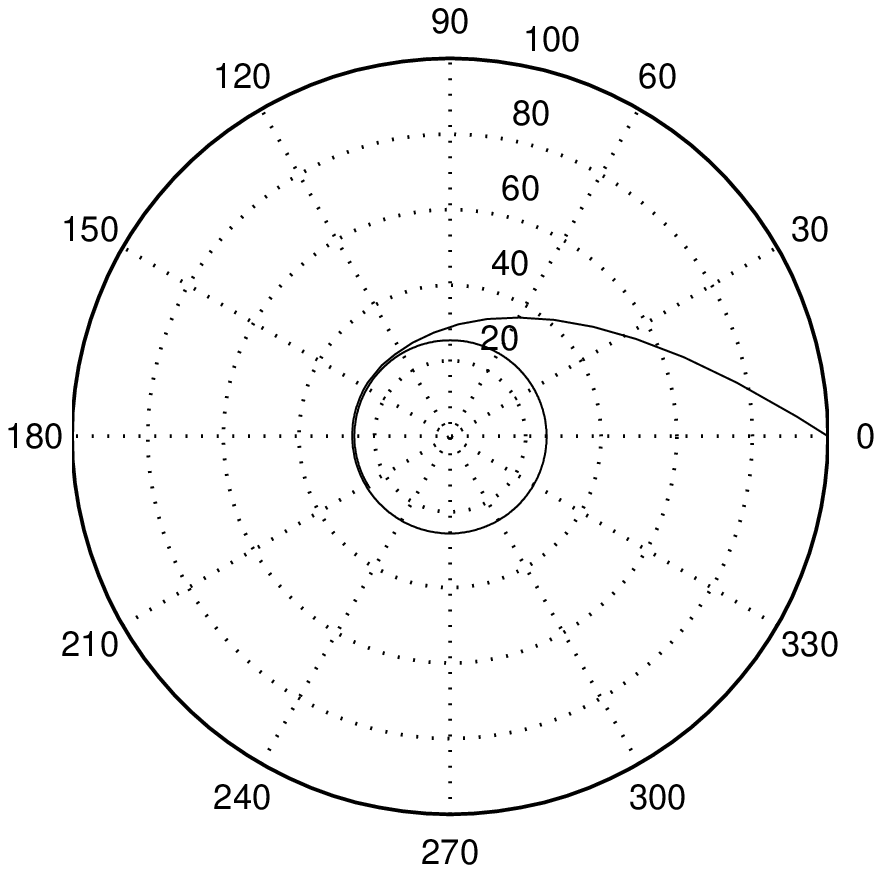}
\caption{\label{fig:gasorbitGas}The orbit of a gas particle free falling to
an attractive point mass from the distance $\bar{r}=100$.}
\end{minipage}
\end{figure}

Figures 3 and 4 show the motion of gas particles at accretion in compare
with free motion of particles in vacuum. It is a good illustration
of fate of gas at accretion for the $\bar{J}=2$. The accreting gas
concentrates at the distances where $\Upsilon(\bar{r})$ has minimum.
For high-temperature gas such places can be very different from this
for free falling particles in vacuum.

\section{Open questions}
Equations considered here are generalizations of equations of motion of a test particle to case 
of small elements of isentropic fluid.

The fact that isentropic fluid can be considered as a curvature of
space-time generates many questions from the point of view of fundamental
physics. The most interesting question is : Is it possible to geometrize fluid without the
limitation "for isentropic fluid"? If - "no", then why not?

\end{document}